\documentclass[letterpaper]{aa}
\usepackage{graphicx}
\usepackage{txfonts}

\begin{document}

   \title{Far-infrared detection of $^{17}$OH towards \object{Sagittarius~B2}
          \thanks{Based on observations with ISO, an ESA project with instruments funded by ESA Member States (especially  the PI countries: France, Germany, the Netherlands and the United Kingdom) with the participation of ISAS and NASA.}
   }

   \author{ E. T. Polehampton
           \inst{1}
           \and
           J. M. Brown
           \inst{2}
           \and
           B. M. Swinyard
           \inst{3}
           \and
           J.-P. Baluteau
           \inst{4}
            }

\offprints{E. Polehampton:  \email{epoleham@mpifr-bonn.mpg.de}}

\institute{Max-Planck-Institut f\"{u}r Radioastronomie, Auf dem H\"{u}gel 69, 53121 Bonn, Germany \\
    \email{epoleham@mpifr-bonn.mpg.de}
\and 
           Physical and Theoretical Chemistry Laboratory, South Parks Road, Oxford, OX1 3QZ, UK \\
    \email{john.m.brown@chem.ox.ac.uk}
\and
           Rutherford Appleton Laboratory, Chilton, Didcot, Oxfordshire, OX11 0QX, UK\\
    \email{B.M.Swinyard@rl.ac.uk}
\and
           Laboratoire d'Astrophysique de Marseille, CNRS \& Universit\'e de Provence, BP 8, F-13376 Marseille Cedex 12, France\\
     \email{Jean-Paul.Baluteau@oamp.fr}
      }

   \date{Received  / accepted }

 \abstract{The frequencies and line strengths of the $^{2}\Pi_{3/2}~J$=5/2--3/2 rotational transition of $^{17}$OH have been calculated from an analysis of its far-infrared laser magnetic resonance spectrum. These results have been used to make the first detection of a pure rotational transition of $^{17}$OH in the ISM. Two resolved components of this transition appear in absorption towards the giant molecular cloud Sagittarius~B2, which was observed at a spectral resolution of 33~km~s$^{-1}$ with the Fabry-P\'{e}rot mode of the ISO Long Wavelength Spectrometer. The corresponding transition of $^{18}$OH was also observed and its line shape was modelled using \ion{H}{i} measurements. The $^{18}$O/$^{17}$O ratio of 3.5 was then used to compare this with the observed $^{17}$OH line shape.

\keywords{Infrared: ISM -- Molecular data -- ISM: molecules -- Line: identification -- ISM: individual objects: Sagittarius~B2  
 }
                 }

   \titlerunning{Far-infrared detection of $^{17}$OH towards \object{Sgr~B2}}
   \maketitle
%
%________________________________________________________________

\section{Introduction}
  
The $^{17}$OH molecule was first observed in the ISM towards the Galactic Centre by Gardner \& Whiteoak (\cite{gardner76}) via its microwave $\Lambda$-doublet transitions. The strongest of these was then observed by Bujarrabal et al. (\cite{bujarrabal}) towards the giant molecular cloud complex, \object{Sagittarius~B2} (Sgr~B2), with similar observations of $^{16}$OH and $^{18}$OH giving the ratio $^{18}\mathrm{OH}/^{17}\mathrm{OH}=3.6\pm0.5$. \object{Sgr~B2} is located $\sim$100~pc from the Galactic Centre and emits a bright background ideal for studying absorption features due to the intervening ISM. This spectrum is dominated by thermal emission from dust and peaks in the far-infrared (FIR) near 80~$\mu$m (see Goicoechea \& Cernicharo \cite{goicoechea1}). The rotational transitions of $^{17}$OH lie in this FIR region and have so far not been observed in the ISM.

We report the first detection of a FIR rotational transition of $^{17}$OH in the ISM. We present accurate frequencies and line strengths for the $^{2}\Pi_{3/2}~J$=5/2--3/2 transition calculated from laboratory measurements. We then present observations of this transition towards \object{Sgr~B2} using the ISO Long Wavelength Spectrometer (LWS; Clegg et al. \cite{clegg}) and model the line shape using the corresponding $^{18}$OH line and previously determined $^{18}$O/$^{17}$O ratio.

\section{Observations and Data Reduction}

\object{Sgr~B2} was observed as part of a wide spectral survey using the ISO LWS Fabry-P\'{e}rot (FP) mode, L03. The whole LWS spectral range, from 47 to 197~$\mu$m, was covered with a spectral resolution of 30--40~km~s$^{-1}$ (see Ceccarelli et al. \cite{ceccarelli} for the first description of this survey). The lowest energy rotational transitions of OH occur in the region 119--120~$\mu$m. At the resolution of the LWS these transitions have two resolved components due to the $\Lambda$-type parity doubling of each rotational level. 

The LWS beam had an effective diameter of 78$\arcsec$ at 119~$\mu$m (Gry et al. \cite{gry}) and was centred at coordinates $\alpha=17^{\mathrm{h}}47^{\mathrm{m}}21.75^{\mathrm{s}}$, $\delta=-28\degr 23\arcmin 14.1\arcsec$ (J2000). This gave the beam centre an offset of 21.5$\arcsec$ from the main FIR peak, which occurs near the radio and mm source, \object{Sgr~B2}~(M) (Goldsmith et al. \cite{goldsmithb}). This pointing was used to exclude the source \object{Sgr~B2}~(N) from the beam. The LWS spectral resolution element at 119~$\mu$m was $\approx$33~km~s$^{-1}$.

The spectral region 119--120~$\mu$m was covered in two separate observations using the LWS detector LW2 (ISO TDT numbers 50601013 and 50700610) with one further observation covering wavelengths below 119.74~$\mu$m (50800416). Each one had a spectral sampling interval of 1/4 resolution element and 3 repeated scans per point. The three observations were processed using the LWS pipeline version 8 and interactively calibrated using routines that appeared as part of the LWS Interactive Analysis (LIA) package version 10. The dark signal, including straylight, was determined as described in Polehampton et al. (\cite{polehampton1}).

Each observation consisted of a series of FP scans at consecutive grating angles, known as ``mini-scans''. Due to uncertainty in the commanded grating angle, the grating response profile must be removed from each mini-scan interactively in order to recover the true spectral shape (see Gry et al. \cite{gry}). This is particularly important for wide lines that may cross one or more mini-scans as it affects the stitching across the joins. For weak lines the overlap region can cause extra uncertainty in the precise line shape if placed unfavourably. A shift for each mini-scan was carefully determined using the standard LIA tool for interactive processing of FP data, FP\_PROC. Accurate grating profile shapes were used - these were recovered from other observations in the \object{Sgr~B2} dataset and corrected for contamination from adjacent FP orders as described in Polehampton et al. (\cite{polehampton2}). Glitches were removed from each observation scan by scan. 

The continuum around the lines was fitted with a polynomial baseline which was then divided into the data. This effectively bypassed the large systematic uncertainty in absolute flux level (caused by multiplicative calibration steps - see Swinyard et al. \cite{swinyard}). The dark signal was accurately determined to be a small fraction of the continuum photocurrent and so the final error in relative line depth is dominated by the statistical noise in the data. To increase the signal to noise ratio, the three observations were corrected to the local standard of rest and co-added. The remaining uncertainty in wavelength is less than 0.004~$\mu$m (or 11~km~s$^{-1}$), corresponding to the error in absolute wavelength calibration (see Gry et al. \cite{gry}). A more detailed description of the calibration and reduction method is given in Polehampton (\cite{polehampton3}).

\section{Laboratory Measurements}

\begin{table}[!t]
\caption{Calculated hyperfine transition frequencies for $J$=5/2--3/2 in the $^{2}\Pi_{3/2}$ state of $^{17}$OH. Errors in the last quoted significant figure are shown in parenthesis. An extended version of this table for higher transitions up to and including the $^{2}\Pi_{3/2}~J$=7/2 level is available electronically at the Centre de Donn\'{e}es Astronomiques de Strasbourg (CDS).
\label{freqs}}
\leavevmode
\footnotesize
\begin{center}
\begin{tabular}[!t]{cccc}
\hline
\hline
Transition       & Frequency & Vacuum     & Line      \\
$F_{i}$--$F_{j}$ &  (MHz)    & Wavelength & Strength  \\
                 &           &  ($\mu$m)  & $S_{ij}$     \\
\hline
0$^{-}$--1$^{+}$ & 2506086 (2) & 119.6257 (1) & 0.2696 \\
1$^{-}$--1$^{+}$ & 2506027 (2) & 119.6286 (1) & 0.5674 \\
2$^{-}$--1$^{+}$ & 2505907 (2) & 119.6343 (1) & 0.3786 \\
1$^{-}$--2$^{+}$ & 2506267 (2) & 119.6171 (1) & 0.2428 \\
2$^{-}$--2$^{+}$ & 2506147 (2) & 119.6228 (1) & 0.8097 \\
3$^{-}$--2$^{+}$ & 2505967 (2) & 119.6314 (1) & 0.9720 \\
2$^{-}$--3$^{+}$ & 2506508 (2) & 119.6056 (1) & 0.1619 \\
3$^{-}$--3$^{+}$ & 2506328 (2) & 119.6142 (1) & 0.8500 \\
4$^{-}$--3$^{+}$ & 2506087 (2) & 119.6257 (1) & 1.822  \\
3$^{-}$--4$^{+}$ & 2506812 (2) & 119.5911 (1) & 0.06748\\
4$^{-}$--4$^{+}$ & 2506570 (2) & 119.6027 (1) & 0.6070 \\
5$^{-}$--4$^{+}$ & 2506266 (2) & 119.6172 (1) & 2.968  \\
\hline
0$^{+}$--1$^{-}$ & 2501707 (2) & 119.8351 (1) & 0.2702 \\
1$^{+}$--1$^{-}$ & 2501656 (2) & 119.8376 (1) & 0.5664 \\
2$^{+}$--1$^{-}$ & 2501555 (2) & 119.8424 (1) & 0.3775 \\
1$^{+}$--2$^{-}$ & 2501884 (2) & 119.8267 (1) & 0.2429 \\
2$^{+}$--2$^{-}$ & 2501783 (2) & 119.8315 (1) & 0.8096 \\
3$^{+}$--2$^{-}$ & 2501632 (2) & 119.8388 (1) & 0.9715 \\
2$^{+}$--3$^{-}$ & 2502124 (2) & 119.8152 (1) & 0.1618 \\
3$^{+}$--3$^{-}$ & 2501973 (2) & 119.8224 (1) & 0.8501 \\
4$^{+}$--3$^{-}$ & 2501773 (2) & 119.8320 (1) & 1.822  \\
3$^{+}$--4$^{-}$ & 2502427 (2) & 119.8007 (1) & 0.06737\\
4$^{+}$--4$^{-}$ & 2502227 (2) & 119.8103 (1) & 0.6071 \\
5$^{+}$--4$^{-}$ & 2501979 (2) & 119.8222 (1) & 2.968 \\ 
\hline
\end{tabular} 
\end{center}
\end{table}

The rotational spectrum of $^{17}$OH is more complicated than the corresponding spectrum of $^{16}$OH and $^{18}$OH as there are up to 6 hyperfine levels associated with the spin of the $^{17}$O nucleus ($I$=5/2), each of which is split into two by the $^{1}$H hyperfine interaction. 

The FIR spectrum of the $^{17}$OH radical has been measured by laser magnetic resonance (LMR) by Leopold et al. (\cite{leopold}). This method involved using a variable magnetic field to tune the frequency of the molecular transition into coincidence with a fixed frequency laser. This meant that zero field frequencies were not measured directly in the experiment and were not given by Leopold et al. (\cite{leopold}).

In order to predict the relevant frequencies and line strengths, we have used a computer program to extrapolate to zero magnetic field. This uses the molecular parameters derived from the LMR experiment and is described in detail in Brown et al. (\cite{brown_a}). This calculation was performed taking account of the $^{17}$O spin. The spin of $^{1}$H causes a further splitting of each component into a closely spaced doublet but this was ignored in the calculation as the doublet separation is well below the resolution limit of the LWS FP and does not affect the final line shape. The zero field line frequencies, wavelengths and line strengths for each hyperfine component of the $J$=5/2--3/2 transition are shown in Table~\ref{freqs}. The line frequencies obtained using this technique are accurate to 2~MHz for lines directly studied in the LMR experiment. The Einstein coefficient for spontaneous emission can be calculated from the line strength as described by Brown et al. (\cite{brown_b}).

\section{Results}

\begin{figure}
\resizebox{\hsize}{!}{\includegraphics{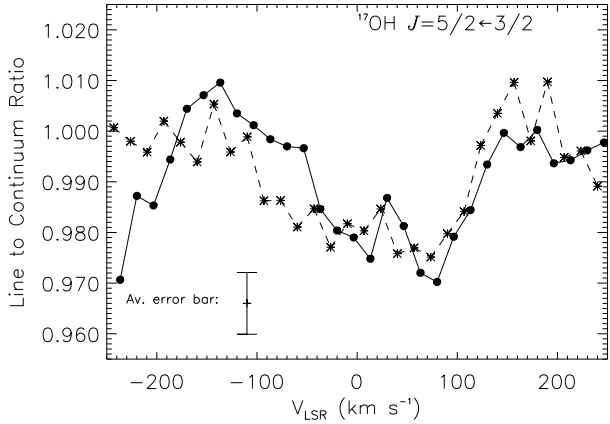}}
\caption{Combined data from three L03 observations showing the two detected features assigned to $^{17}$OH. The velocity scale is relative to the strongest components in Table~1: 119.6172~$\mu$m (solid line) and 119.8222~$\mu$m (dashed line). The data were binned at 1/2 spectral resolution element and the average error in each bin across the line is shown. The sharp drop on the left hand side of the 119.6~$\mu$m line is due to the corresponding transition in $^{16}$OH.}
\label{2comp}
\end{figure}

Two absorption lines are present in the spectrum of \object{Sgr~B2} which agree well with the calculated wavelengths (Fig.~\ref{2comp}) and we assign them to the two $\Lambda$-doubling components of the $J$=5/2--3/2 transition. We are confident that these features are real as they occur in all three independent observations. Furthermore, the two components agree well in shape as expected from all other observed transitions of OH towards this source (see Goicoechea \& Cernicharo \cite{goicoechea2}). They cannot be due to ``ghost lines'' (caused by contamination from adjacent FP orders; see Gry et al. \cite{gry}) because there are no strong lines at the wavelengths of neighbouring orders. There is strong $^{16}$OH absorption next to the detected features but this has a separation much less than the offset of nearby FP orders.

The features show a minimum near the expected velocity of \object{Sgr~B2}~(M) ($\approx+65$~km~s$^{-1}$; e.g. Mart\'{\i}n-Pintado et al. \cite{martinpintado}) with additional absorption at negative velocities due to absorbing clouds corresponding to the main galactic spiral arms crossing the line of sight (e.g. Greaves \& Williams \cite{greaves94}). These clouds have a low density and temperature (Greaves \cite{greaves}) and are illuminated by the galactic interstellar radiation field.

The lines of $^{16}$OH and $^{18}$OH originating in the ground rotational state also show this broad structure, although with a clearer minimum at the velocity of \object{Sgr~B2} (see Goicoechea \& Cernicharo \cite{goicoechea2}). The two components of the $J$=5/2--3/2 transition in $^{18}$OH occur at wavelengths of 119.9651~$\mu$m and 120.1718~$\mu$m (Morino et al. \cite{morino}) and show good agreement in shape and depth in the \object{Sgr~B2} spectrum.

\section{Line Modelling}

\begin{figure}
\resizebox{\hsize}{!}{\includegraphics{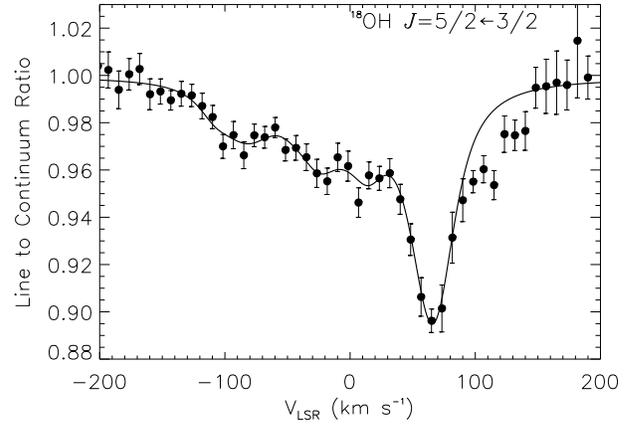}}
\caption{$^{18}$OH absorption line data after co-adding the two $\Lambda$-doublet components. The best fit using a model derived from \ion{H}{i} observations is shown. The data were binned at 1/4 spectral resolution element.}
\label{18oh}
\end{figure}

Due to the large errors when directly fitting the weak $^{17}$OH absorption, the $^{18}$OH line was modelled and used with the $^{18}$O/$^{17}$O ratio to predict the $^{17}$OH shape. 

Models of photo-dissociation regions show that OH has highest concentration at the interface between \ion{H}{i} and H$_{2}$ in dense clouds illuminated by UV radiation (Sternberg \& Dalgarno \cite{sternberg}). We assume that this is also true in the low density clouds towards \object{Sgr~B2} and used \ion{H}{i} to trace the line of sight features. High resolution measurements of the \ion{H}{i} 21~cm line have been carried out towards \object{Sgr~B2}~(M) by Garwood \& Dickey (\cite{garwood}). Ten velocity features are present in their spectrum and we fixed the line widths and velocities to be the same for $^{18}$OH. The optical depth in each component was adjusted and the spectrum convolved to the resolution of the LWS FP. The best fit was found by minimising $\chi^2$.

The column density of $^{18}$OH in the ground state for each component was calculated from the optical depths and lines widths. This is a good measure of the total column density in the line of sight clouds where only ground state transitions have been observed (Goicoechea \& Cernicharo \cite{goicoechea2}). The column density in the fit summed over the whole line of sight is $(1.8\pm0.2)\times10^{14}$~cm$^{-2}$, marginally consistent with the total column density found from KAO measurements which gave $N(\mathrm{^{18}OH})\geq2\times10^{14}$~cm$^{-2}$ (Lugten et al. \cite{lugten}). 

In the \object{Sgr~B2} envelope, $^{16}$OH transitions involving higher levels up to 420~K have been observed and radiative transfer modelling by Goicoechea \& Cernicharo (\cite{goicoechea2}) gave $N(^{18}\mathrm{OH})=(6\pm2)\times10^{13}$~cm$^{-2}$. The column density derived from components associated with \object{Sgr~B2} in the $^{18}$OH fit was $(9.8\pm1.6)\times10^{13}$~cm$^{-2}$, close to the radiative transfer value. The model, results and associated errors and assumptions for $^{18}$OH have been described in detail by Polehampton (\cite{polehampton3}) and will be presented in a forthcoming paper.

In order to predict the shape of the $^{17}$OH absorption from the best fitting model of $^{18}$OH, the $^{18}$O/$^{17}$O ratio was used. This has been measured near the Galactic Centre using several molecular tracers and these show good consistency (Gardner \& Whiteoak \cite{gardner76}; Wannier et al. \cite{wannier}; Penzias \cite{penzias}; Gu\'{e}lin et al. \cite{guelin}; Bujarrabal et al. \cite{bujarrabal}). The weighted average of these values gives $^{18}$O/$^{17}$O=3.5$\pm$0.9, in agreement with $^{18}$O/$^{17}$O measured throughout the Galactic Disk by Penzias (\cite{penzias}). However, this is lower than the standard Solar System ratio ($5.237\pm0.004$; Baertschi \cite{baertschi}; Fahey et al. \cite{fahey}). This discrepancy is a long standing problem (e.g. see Prantzos et al. \cite{prantzos}).

\begin{figure}[!t]
\resizebox{\hsize}{!}{\includegraphics{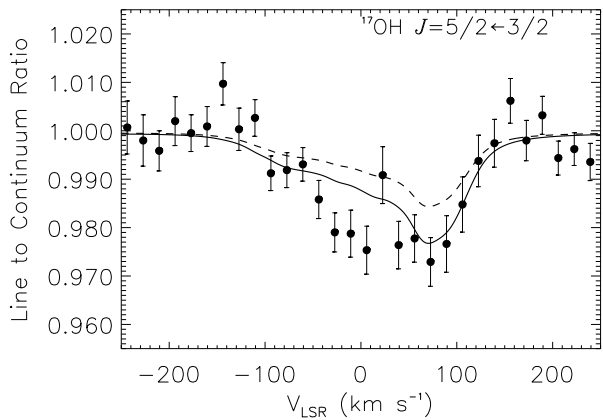}}
\caption{$^{17}$OH absorption line data with model derived from the observed $^{18}$OH absorption and $^{18}$O/$^{17}$O ratio of 3.5 (solid line). The data were binned at 1/2 spectral resolution element. The predicted shape using the Solar System $^{18}$O/$^{17}$O ratio of 5.2 is also shown for comparison (dashed line).}
\label{17oh_fit}
\end{figure}

As the ratio between $^{18}$O and $^{17}$O is not large and both OH lines are optically thin, opacity and excitation effects should not affect the final ratio. The $^{17}$OH column density for each line of sight feature was calculated by taking the values from the $^{18}$OH fit, dividing by 3.5 and splitting up between the hyperfine states within each rotational level (according to the relative line strengths of their transitions given in Table~\ref{freqs}). This allowed the calculation of the optical depth for each hyperfine transition and when combined with the line widths and velocities from \ion{H}{i} gave a high resolution model of the $^{17}$OH absorption. The shape has a more complicated structure than for $^{18}$OH as the hyperfine components have a comparable separation to the line of sight features. Figure~\ref{17oh_fit} shows this model (after convolution with the FP response profile) with the co-added $^{17}$OH data. The large number of hyperfine components cause a significant modification of the line shape in the convolved spectrum with a reduction in the relative absorption of the 65~km~s$^{-1}$ feature compared with $^{18}$OH. Figure~\ref{17oh_fit} also clearly shows that there is too little absorption when the Solar System ratio of 5.2 is used.

The model fits the $^{17}$OH data well in the region of \object{Sgr~B2} but there is extra absorption in the the data near 0~km~s$^{-1}$. This may be a real effect as the two resolved $\Lambda$-doublet components agree very well in shape. It indicates that the $^{18}$O/$^{17}$O ratio should be lower than 3.5 in the line of sight clouds, although this seems unlikely given the measurements by Penzias (\cite{penzias}) that gave a constant ratio throughout the Galactic Disk. It is also unlikely that the extra absorption is due to transitions from different species as there would have to be two lines with the same relative spacing from each $^{17}$OH component.

An instrumental effect that could cause this distortion is the mini-scan overlap regions discussed in Sect.~2. Both $^{17}$OH components occurred in the same position relative to the join between mini-scans and so the same effect could have been reproduced in each one. Further observations of these lines will be necessary to confirm the shape. The accurate wavelengths and line strengths presented here will allow this line to be searched for with future instrumentation such as that on SOFIA. 
 
\begin{acknowledgements}

We thank the referee for useful comments and suggestions.

\end{acknowledgements}


\begin{thebibliography}{}

\bibitem[1976]{baertschi}
Baertschi, P. 1976, Earth \& Planetary Sci. Lett., 31, 341
\bibitem[1978]{brown_a}
Brown, J. M., Kaise, M., Kerr, C. M. L., \& Milton, D. J. 1978, Mol. Phys., 36, 553
\bibitem[1982]{brown_b}
Brown, J. M., Schubert, J. E., Evenson, K. M., \& Radford, H. E. 1982, ApJ, 258, 899
\bibitem[1983]{bujarrabal}
Bujarrabal, V., Cernicharo, J., \& Gu\'{e}lin, M. 1983, A\&A, 128, 355
\bibitem[2002]{ceccarelli}
Ceccarelli, C., Baluteau, J.-P., Walmsley, M., et al. 2002, A\&A, 383, 603 
\bibitem[1996]{clegg}
Clegg, P. E., Ade, P. A. R., Armand, C., et al. 1996, A\&A, 315, L38
\bibitem[1987]{fahey}
Fahey, A. J., Goswami, J. N., McKeegan, K. D., \& Zinner, E. K. 1987, ApJ, 323, L91
\bibitem[1976]{gardner76}
Gardner, F. F., \& Whiteoak, J. B. 1976, MNRAS, 176, 57
\bibitem[1989]{garwood}
Garwood, R. W., \& Dickey, J. M. 1989, ApJ, 338, 841
\bibitem[2001]{goicoechea1}
Goicoechea, J. R., \& Cernicharo, J. 2001, ApJ, 554, L213
\bibitem[2002]{goicoechea2}
Goicoechea, J. R., \& Cernicharo, J. 2002, ApJ, 576, L77
\bibitem[1992]{goldsmithb}
Goldsmith, P. F., Lis, D. C., Lester, D. F., \& Harvey, P. M. 1992, ApJ, 389, 338
\bibitem[1994]{greaves94}
Greaves, J. S., \& Williams, P. G. 1994, A\&A, 290, 259
\bibitem[1995]{greaves}
Greaves, J. S. 1995, MNRAS, 273, 918 
\bibitem[2002]{gry}
Gry, C., Swinyard, B., Harwood, A., et al. 2002, ISO Handbook Volume III (LWS), ESA SAI-99-077/Dc
\bibitem[1982]{guelin}
Gu\'{e}lin, M., Cernicharo, J., \& Linke, R. A. 1982, ApJ, 263, L89
\bibitem[1987]{leopold}
Leopold, K. R., Evenson, K. M., Comben, E. R., \& Brown, J. M. 1987, J. Mol. Spectrosc., 122, 440
\bibitem[1986]{lugten}
Lugten, J. B., Stacey, G. J., \& Genzel, R. 1986, BAAS, 18, 1007
\bibitem[1990]{martinpintado}
Mart\'{\i}n-Pintado, J., de Vicente, P., Wilson, T. L., \& Johnston, K. J. 1990, A\&A, 236, 193
\bibitem[1995]{morino}
Morino, I., Odashima, H., Matsushima, F., Tsunekawa, S., \& Takagi, K. 1995, ApJ, 442, 907
\bibitem[1981]{penzias}
Penzias, A. A. 1981, ApJ, 249, 518
\bibitem[2002]{polehampton3}
Polehampton, E. T. 2002, Ph.D. Thesis (Oxford University)
\bibitem[2002]{polehampton1}
Polehampton, E. T., Baluteau, J.-P., Ceccarelli, C., Swinyard, B. M., \& Caux, E. 2002, A\&A, 388, L44
\bibitem[2001]{polehampton2}
Polehampton, E. T., Swinyard, B. M., Sidher, S. D., \& Baluteau, J.-P. 2001, in The Calibration Legacy of the ISO Mission, ESA SP-481, in press
\bibitem[1996]{prantzos}
Prantzos, N., Aubert, O., \& Audouze, J. 1996, A\&A, 309, 760
\bibitem[1995]{sternberg}
Sternberg, A., \& Dalgarno, A. 1995, ApJS, 99, 565
\bibitem[1998]{swinyard}
Swinyard, B. M., Burgdorf, M. J., Clegg, P. E., et al. 1998, SPIE, 3354, 888
\bibitem[1976]{wannier}
Wannier, P. G., Lucas, R., Linke, R. A., et al. 1976, ApJ, 205, L169

\end{thebibliography}
\end{document}